\documentclass[twocolumn,showpacs,preprintnumbers,amsmath,amssymb,prb]{revtex4}

\usepackage{graphicx}
\usepackage{dcolumn}
\usepackage{bm}

\begin{document}


\title{\textit{Ab initio} electronic structure calculation of hollandite vanadate K$_2$V$_8$O$_{16}$}

\author{M. Sakamaki$^1$}
\author{S. Horiuchi$^2$}
\author{T. Konishi$^1$}
\author{Y. Ohta$^2$}
\affiliation{$^1$Graduate School of Advanced Integration Science, Chiba University, Chiba 263-8522, Japan}
\affiliation{$^2$Department of Physics, Chiba University, Chiba 263-8522, Japan}

\date{\today}

\begin{abstract}
An \textit{ab initio } electronic structure calculation based 
on the generalized gradient approximation in the density 
functional theory is carried out to study the basic electronic 
states of hollandite vanadate K$_2$V$_8$O$_{16}$.  
We find that the states near the Fermi energy consist 
predominantly of the three $t_{2g}$-orbital components and the 
hybridization with oxygen $2p$ orbitals is small.  
The $d_{yz}$ and $d_{zx}$ orbitals are exactly degenerate 
and are lifted from the $d_{xy}$ orbital.  The calculated band 
dispersion and Fermi surface indicate that the system is not 
purely one-dimensional but the coupling between the VO double 
chains is important.  Comparison with available experimental 
data suggests the importance of electron correlations in this 
system.  
\end{abstract}

\pacs{71.30.+h, 71.20.Be, 71.28.+d}

\maketitle

\section{Introduction}

The metal-insulator phase transition associated with charge and 
orbital ordering has been one of the central issues in physics 
of strongly correlated electron systems.  
Recently, Isobe {\it et al.}\cite{isobe} reported that the 
metal-insulator transition occurs in hollandite vanadate 
K$_2$V$_8$O$_{16}$ at $\sim$160 K, which is accompanied by a rapid 
reduction of the magnetic susceptibility.  A characteristic 
superlattice of $\sqrt{2}a\times \sqrt{2}a\times 2c$ is observed 
below the transition temperature.  A possible charge-ordering 
phase transition accompanied by the spin-singlet formation has 
thereby been suggested.\cite{isobe}  Quite recently, the phase 
diagram under high pressures has also been obtained,\cite{yamauchi} 
where a variety of charge-ordered phases have been suggested 
to appear.  The metal-insulator transition has also been observed 
in Bi$_x$V$_8$O$_{16}$\cite{waki,shibata} and 
K$_2$Cr$_8$O$_{16}$.\cite{hasegawa} 

The crystal structure of K$_2$V$_8$O$_{16}$ belongs to a group 
of hollandite-type phases and has a V$_{8}$O$_{16}$ framework 
composed of double strings of edge-shared VO$_6$ octahedra.  
The system may be regarded as a one-dimensional version of 
LiVO$_2$ known as a possible orbital-ordering system of the 
$t_{2g}$ orbitals on the triangular lattice of $S=1$ spins.\cite{pen}  
The present system K$_2$V$_8$O$_{16}$ has the average valence 
of V$^{3.75+}$ and thus is in the mixed valent state of 
V$^{3+}$:~V$^{4+}=3d^2:3d^1=1:3$.  Thus, the central issue 
in the present system is the mechanism of the metal-insulator 
transition concerning how the highly frustrated spin, charge, 
and orbital degrees of freedom at high temperatures are relaxed 
by lowering temperatures and what type of order is realized 
in the ground state.  

In our previous paper,\cite{horiuchi} we studied this material 
in the strong correlation limit; we set up a high-energy model 
Hamiltonian and applied a strong-coupling perturbation theory 
to obtain the low-energy effective spin-orbit Hamiltonian.  
The obtained effective Hamiltonian was analyzed numerically and 
the possible orbital and spin structure of the ground state of 
the system was proposed.\cite{horiuchi}

However, so far not much is known even for the basic electronic 
states of the system, such as the band dispersion, Fermi surface, 
density of states, etc., both experimentally and theoretically.  
The purpose of the present paper is therefore to present 
an \textit{ab initio} electronic structure calculation of 
this system to clarify its basic electronic structure.  
Here, we use the generalized gradient approximation (GGA) 
for the electron correlations in the density functional theory.  
The study is thus in the weak correlation limit but we hope 
that one can learn much of the basic electronic states of 
this material.  

In this paper, we will show the following: 
The electronic states near the Fermi level are dominated 
by the three $t_{2g}$-orbital components, where the hybridization 
with the oxygen $2p$ orbitals is small.  The $d_{yz}$ and 
$d_{zx}$ orbitals are exactly degenerate and are lifted 
from the $d_{xy}$ orbital (hereafter we use the notation of 
the axes $x$, $y$, and $z$ defined in Ref.,~\cite{horiuchi} 
also see Fig.~5 below).  
The calculated band structure and Fermi surface indicate 
that the system is not purely one-dimensional but the 
coupling between the VO double chains is important.  
We find no nesting features in the Fermi surfaces that 
contribute to the instability of the doubling of the 
unit cell along the $c$ axis, suggesting the observed lattice 
instability to be the strong-coupling origin.  
We will compare our calculated results with experiment 
such as photoemission spectroscopy\cite{PES} and nuclear 
magnetic resonance (NMR),\cite{NMR} from which we will 
discuss the implications on the electronic states and 
peculiarity of this material.  

We hope that the present study, if combined with the 
strong-coupling approach, will give a reliable theory 
to explain the mechanism of the electronic phase transition 
of this intriguing material.

\section{Method of calculation}

We here employ the computer code WIEN2k\cite{wien2k} 
based on the full-potential linearized 
augmented-plane-wave (FLAPW) method.  
We have tested both the local density approximation (LDA) 
and generalized gradient approximation (GGA) for the 
exchange-correlation potential\cite{PW92,PBE96} in the density 
functional theory.  In this paper, we will however present 
only the results of GGA because no significant differences 
are found between the results of LDA and GGA.  
The spin-orbit interaction is not taken into account.  
No spin-polarization is assumed.  
In the self-consistent calculations, we use 4,335 
${\bm k}$ points in the irreducible part of the Brillouin 
zone (see Fig.~1) with an anisotropic sampling in order 
to achieve better convergence.  We use the plane-wave cutoff 
of $K_{\rm max}=4.24$ Bohr$^{-1}$.  

We assume the experimental crystal structure of 
K$_2$V$_8$O$_{16}$ observed at room temperature (above 
the metal-insulator transition) with the lattice constants 
of $a=9.963$ and $c=2.916$ \AA.\cite{abriel}  
The Bravais lattice is body-centered tetragonal (see Fig.~1) 
and the primitive unit cell contains four V ions, 
one K ion, and eight O ions.  
We use the code XCrySDen\cite{kokalj} for graphical purposes.  

\begin{figure}[thb]
\begin{center}
\resizebox{8.5cm}{!}{\includegraphics{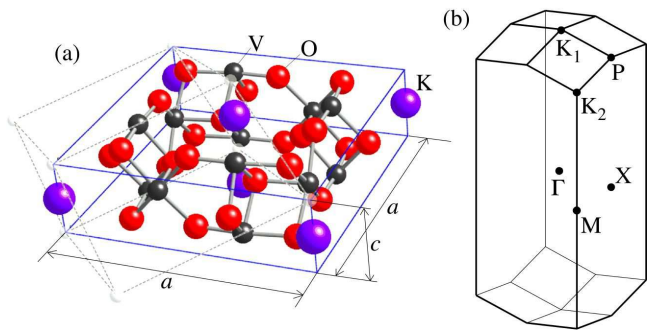}}\\
\caption{(Color online) Schematic representation of 
(a) the unit cell of the body-centered tetragonal lattice 
(solid lines) and (b) Brillouin zone of K$_2$V$_8$O$_{16}$.  
In (a), the primitive unit cell is also shown in the 
thin dotted lines.  In (b), the symbols represent 
$\Gamma(0,0,0)$, M$(2\pi/a,0,0)$, X$(\pi/a,\pi/a,0)$, 
P$(\pi/a,\pi/a,\pi/c)$, K$_1$$(0,0,\pi(1/c+c/a^2))$, 
and K$_2$$(2\pi/a,0,\pi(1/c-c/a^2))$, where K$_1$ and 
K$_2$ are equivalent.}
\label{fig.1}
\end{center}
\end{figure}

\section{Results of calculation}

\subsection{Band dispersion}

The calculated band dispersion near the Fermi energy 
is shown in Fig.~2.  
We find that there are 12 $t_{2g}$ bands of the V 
$3d$ orbitals, 5 of which cross the Fermi level.   
We find the highly dispersive bands along the 
$\Gamma$-K$_1$, X-P, and M-K$_2$ lines 
and weakly dispersive bands along the $\Gamma$-X, 
P-K$_1$, and K$_1$-K$_2$ lines, reflecting the 
one-dimensionality of the electronic state.  
However, the band structure is not simple even near 
the Fermi energy, suggesting that the contributions 
from the three-dimensionality are not negligible; 
in particular, the relatively large dispersion along 
the $\Gamma$-M line reflects the coupling between 
the VO double chains.  
Thus, to consider the low-energy electronic properties 
of this system, the three-dimensionality or the 
coupling between the VO double chains is essential.  
This result is consistent with the observed rather 
small anisotropy of the electric resistivity of the 
single crystal of this material.\cite{yamauchi}

\begin{figure}[thb]
\begin{center}
\resizebox{7.5cm}{!}{\includegraphics{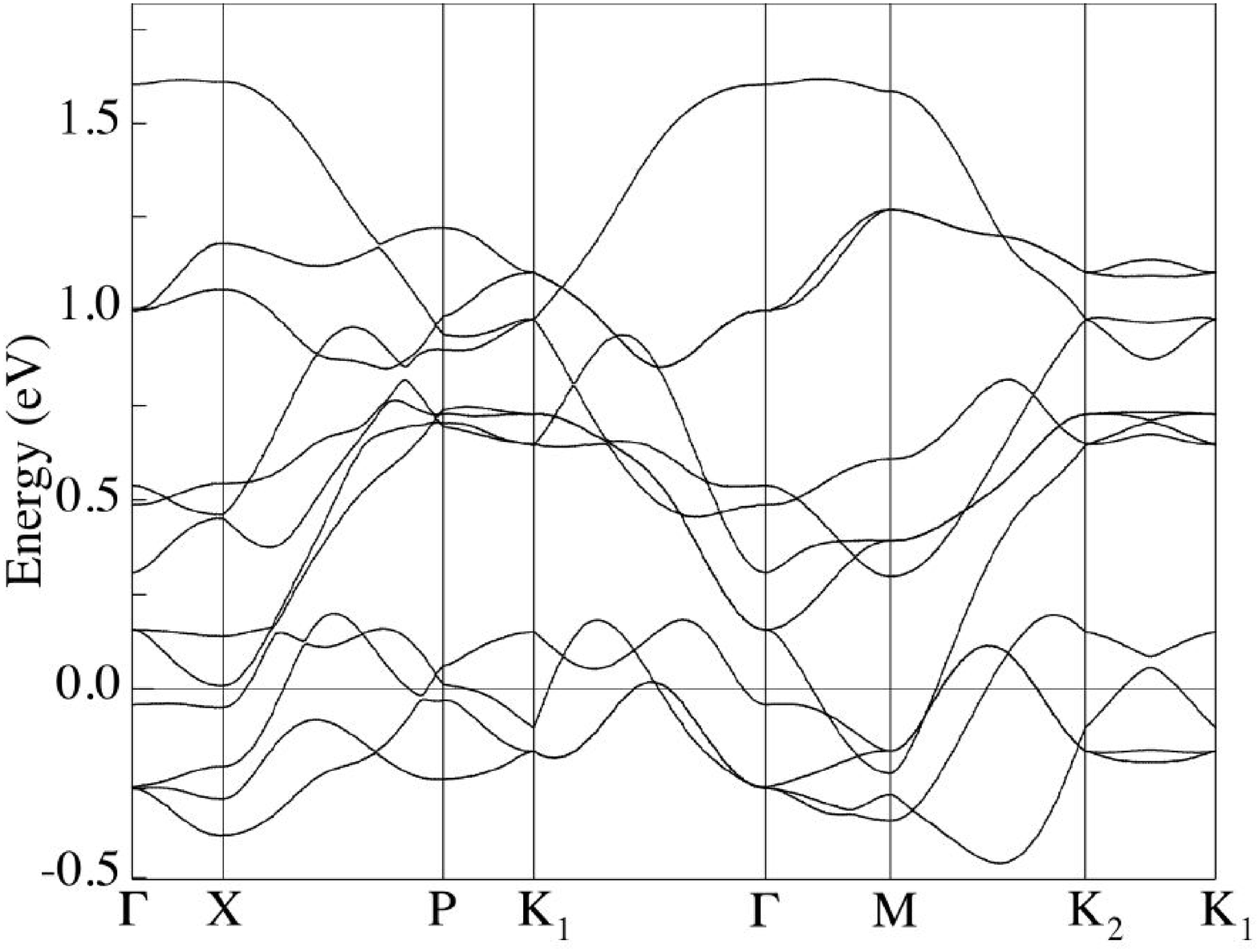}}\\
\caption{Calculated band dispersion of K$_2$V$_8$O$_{16}$ 
near the Fermi level (horizontal line).  
There are 12 $t_{2g}$ bands of the V $3d$ orbitals, 
5 of which cross the Fermi level.  
The labels of the ${\bm k}$-points are shown 
in Fig.~1.}
\label{fig.2}
\end{center}
\end{figure}

\subsection{Density of states}

The calculated density of states is shown in Fig.~3, both 
in a wide energy range in (a) and near the Fermi energy in (b).  
We find that the V $3d$ band is located between $-0.5$ eV 
and $4.5$ eV and is well separated from the O $2p$ band 
located between $-7.5$ eV and $-2.5$ eV.  
The hybridization between V $3d$ and O $2p$ orbitals seems to be 
rather small, justifying the use of the strong-coupling model 
with only the $d$ orbitals.\cite{horiuchi}  
We also find that the $t_{2g}$ band is well separated from the 
$e_g$ band in the V $3d$ bands, corresponding to a large value 
of the crystal field splitting of $10D_q\simeq 2$ eV.  
Thus, the low-energy properties of this system is essentially 
governed by the three $t_{2g}$ orbitals, $d_{xy}$, $d_{yz}$, 
and $d_{zx}$.  

The calculated orbital-decomposed partial densities of states 
$\rho_\alpha(\varepsilon)$ ($\alpha=xy, yz, zx$) are shown in 
Fig.~3(b).  We first find that the total density of states at the 
Fermi level is very high, $30.4$ states/f.u./eV (see Fig.~3(a)), 
which comes mainly from the three $t_{2g}$ orbitals; the large 
ferromagnetic spin fluctuations observed in the NMR 
experiment\cite{NMR} may be related to this high density of 
states at the Fermi level.  
We then find that the three $t_{2g}$ orbitals equally contribute 
to the low-energy states near the Fermi energy.  
More precisely, we note that, even in the presence of the 
distortions of the VO$_6$ octahedra, the relation 
$\rho_{yz}(\varepsilon)=\rho_{zx}(\varepsilon)$ strictly holds.  
This is due to the inversion symmetry along the $c$-axis 
preserved even in the presence of the distortion.\cite{horiuchi}  
We find that the width of the $d_{xy}$ band is slightly smaller 
than the width of the $d_{yz}$ and $d_{zx}$ bands.  
We also find that the number of electrons in the $d_{xy}$ orbital 
is slightly larger than that in the $d_{yz}$ and $d_{zx}$ 
orbitals; in the V atomic sphere, they are $0.345$ and $0.246$, 
respectively, in the energy range of the $t_{2g}$ bands 
(see Sec.~III D).  

\begin{figure}[thb]
\begin{center}
\resizebox{8.2cm}{!}{\includegraphics{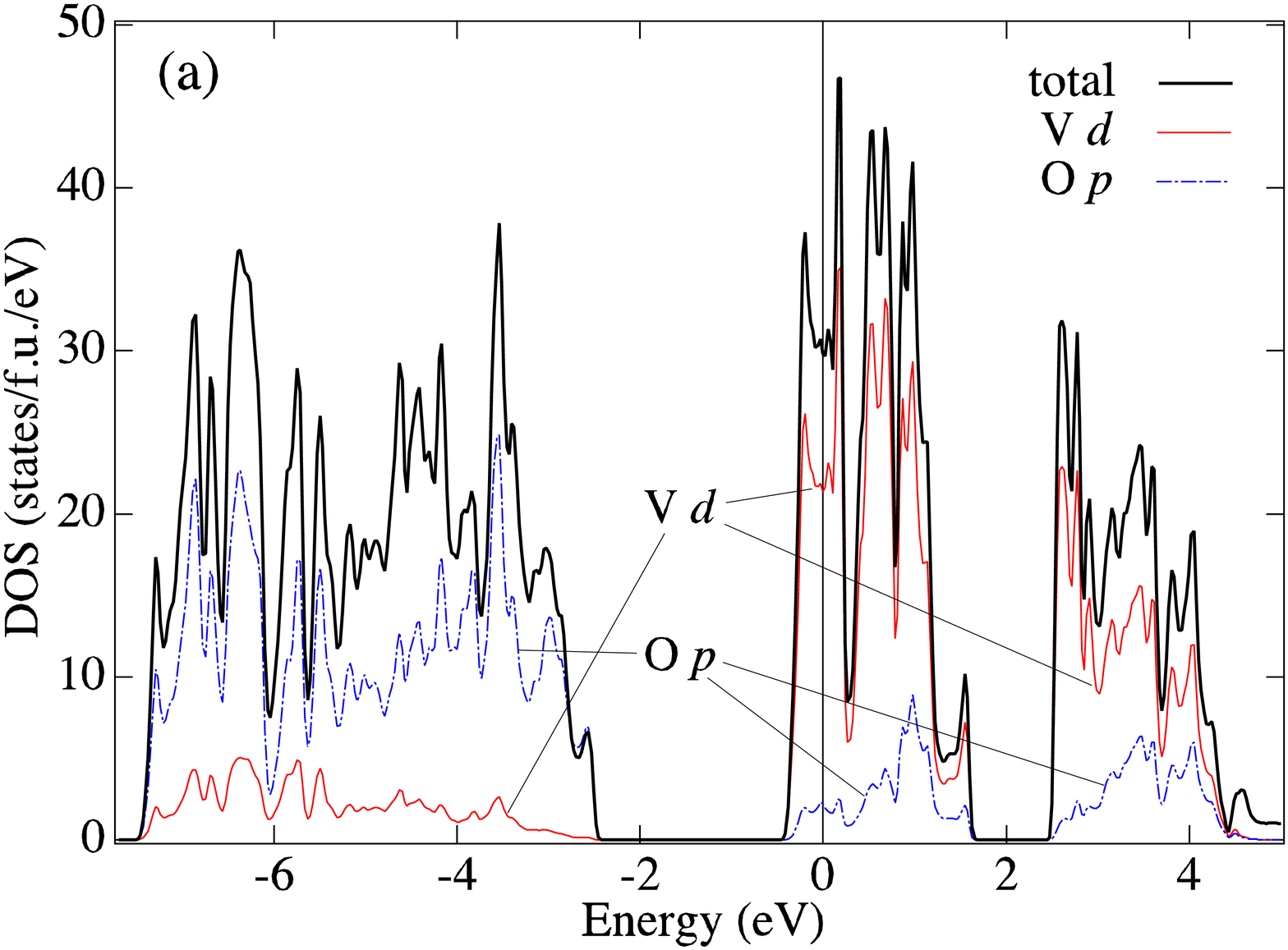}}\\
\resizebox{5.5cm}{!}{\includegraphics{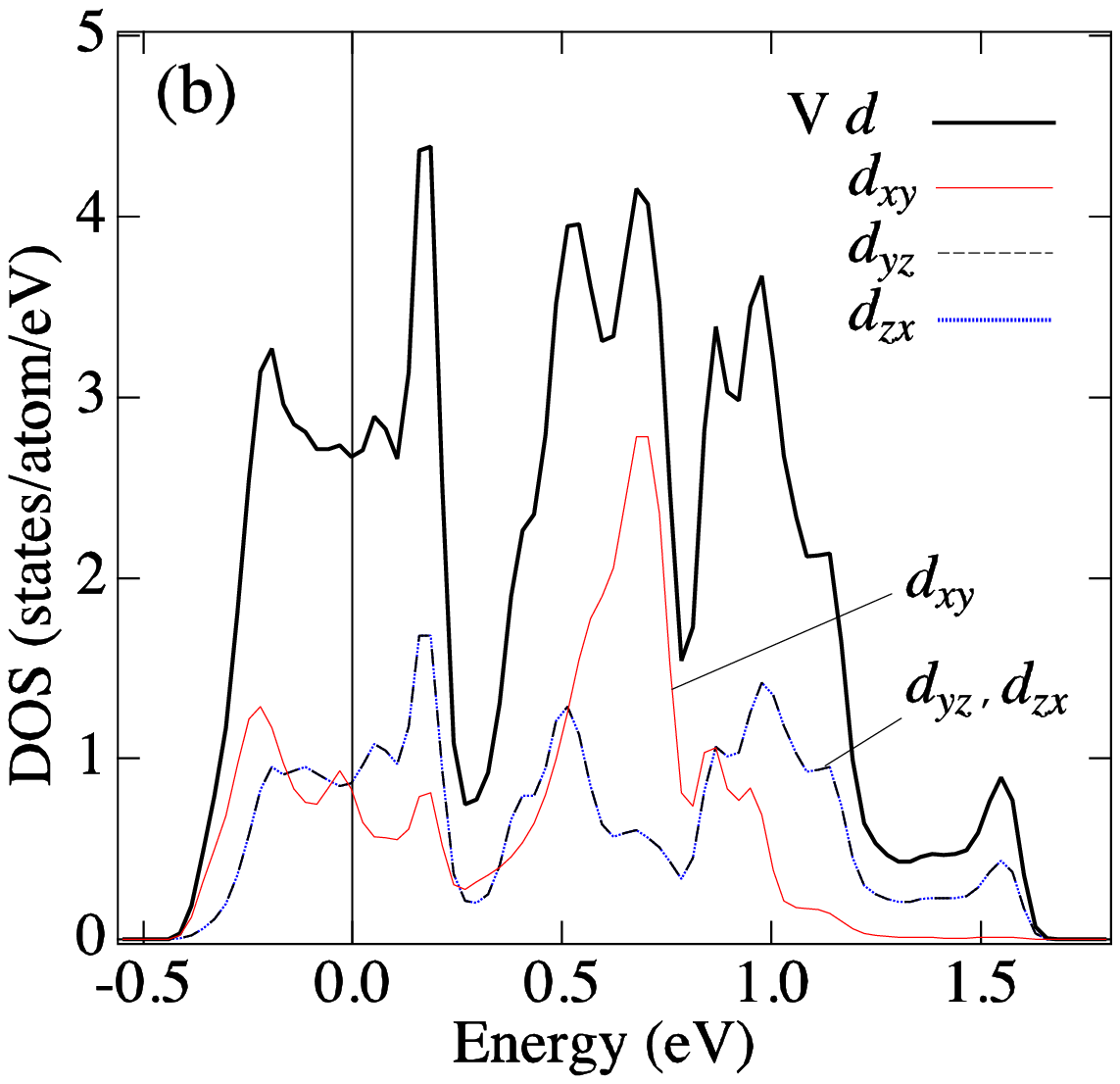}}\\
\caption{(Color online) (a) Calculated density of states 
(per formula unit (f.u.)) of K$_2$V$_8$O$_{16}$ 
in a wide energy range.  
(b) Calculated orbital-decomposed partial density of 
states (per atom) near the Fermi level.  
The Fermi level is indicated by the vertical line.  
Contributions from the $d_{yz}$ and $d_{zx}$ orbitals 
(thin dotted lines) are exactly degenerate.}
\label{fig.3}
\end{center}
\end{figure}

\subsection{Fermi surface}

The calculated results for the Fermi surface are 
shown in Fig.~4.  
There are 12 $t_{2g}$ bands, 5 of which cross the Fermi 
level and form the Fermi surface.  
If the system were strictly one-dimensional, a pair of 
the parallel Fermi surfaces should appear at 
$k_z=\pm k_{\rm F}$.  
We actually find such types of the pairs of the Fermi 
surfaces in Figs.~4 (c) and (d), but the surfaces are 
rather distorted and moreover there are other types of 
the Fermi surfaces present.  In particular, the Fermi 
surface shown in Fig.~4(e) comes from the dispersive 
band along the $\Gamma$-M line of the Brillouin zone 
(see Fig.~2) and is related to the coupling between 
the VO double chains.  Therefore, we may conclude that 
the system is not purely one-dimensional, but the 
three-dimensionality or the coupling between the VO 
double chains is important.  

We should point out that the possible nesting features 
in the two nearly-parallel Fermi surfaces seen in 
Fig.~4(c) does not lead to the $2k_{\rm F}$ instability 
corresponding to the observed\cite{isobe} doubling of 
the unit cell along the $c$ axis expected in the 
weak-coupling theory because the distance between the 
two Fermi surfaces is too short.  No other Fermi 
surfaces are seen to contribute to the instability of 
the $\sqrt{2}a\times\sqrt{2}a\times 2c$ superlattice 
formation.  We therefore consider the observed lattice 
instability to be the strong-coupling origin, such as the 
long-range Coulomb interactions leading to charge 
orderings.\cite{horiuchi}  

\begin{figure}[thb]
\begin{center}
\resizebox{6.3cm}{!}{\includegraphics{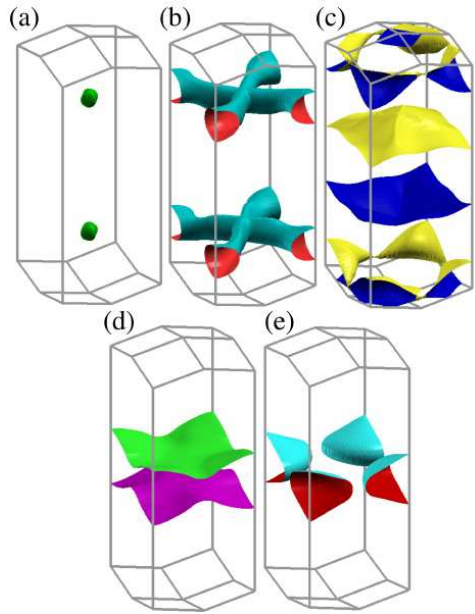}}\\
\caption{(Color online) Calculated Fermi surfaces of 
K$_2$V$_8$O$_{16}$.  The 53rd to 57th bands counted 
from the lowest are shown in (a) to (e), respectively.}
\label{fig.4}
\end{center}
\end{figure}

\subsection{Valence electron distributions}

The spatial distribution of the valence electrons of 
K$_2$V$_8$O$_{16}$ in the energy range between $-0.5$ eV 
and the Fermi level is shown in Fig.~5.  The $t_{2g}$ 
electrons mainly contribute to this distribution.  
We find that the result can be understood if we assume 
that the valence electrons are confined predominantly 
in the ligand field of each VO$_6$ octahedron; i.e., 
the distribution can be projected mainly onto the $d_{xy}$, 
$d_{yz}$, and $d_{zx}$ orbitals.  
More precisely, however, we find that the distribution 
is somewhat compressed in the $z$ direction (toward the 
apical O) and spread in the $xy$ plane, indicating that 
the number of $d_{xy}$ electrons is larger than that of 
$d_{yz}$ and $d_{zx}$ electrons.  
This is consistent with the results for the 
orbital-decomposed partial density of states (see Fig.~3), 
where the area below the Fermi level in the $d_{xy}$ 
component is slightly larger than that in the  $d_{yz}$ 
and $d_{zx}$ components.  We also note that the electron 
distribution reflects the symmetry that the $d_{yz}$ and 
$d_{zx}$ orbitals are equivalent.  
In the atomic limit, this degeneracy corresponds to the 
fact that the atomic energy levels of the $d_{yz}$ and 
$d_{zx}$ orbitals are exactly degenerate in the ionic model 
and are located slightly lower in energy than the atomic level 
of the $d_{xy}$ orbital.\cite{horiuchi}  The orbital 
ordering in this system has thereby been suggested in 
the strong coupling theory.\cite{horiuchi}  

\begin{figure}[thb]
\begin{center}
\resizebox{8.3cm}{!}{\includegraphics{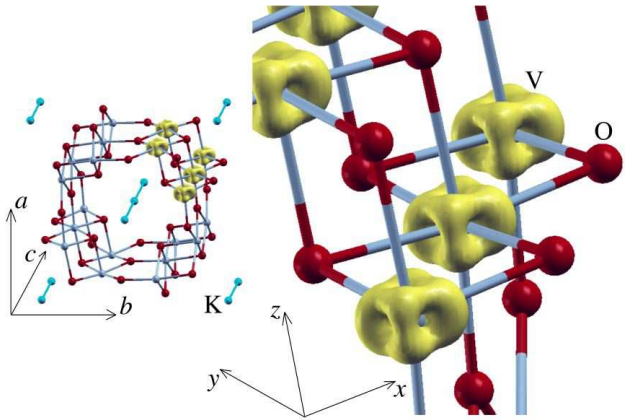}}\\
\caption{(Color online) Calculated spatial density 
distributions of the valence electrons of 
K$_2$V$_8$O$_{16}$ in the energy range between 
$-0.5$ eV and the Fermi level, where the $t_{2g}$ 
electrons mainly contribute.  
The isovalue of 0.08 electrons/Bohr$^3$ is used.}
\label{fig.5}
\end{center}
\end{figure}

\section{Discussion}

Our calculated results presented here contains series of 
theoretical predictions, which should be checked by the 
experimental studies.  

Let us first compare our calculated results with the 
results of the photoemission spectroscopy 
experiment.\cite{PES}  It has been reported that, 
above the transition temperature, 
the coherent quasiparticle peak appears at the binding 
energy of $\varepsilon_b\simeq 0-0.5$ eV (where the Fermi 
energy is set to 0), which we find corresponds well to 
the $t_{2g}$ band below the Fermi energy shown in Fig.~3.  
The large incoherent spectral weight also appears at the binding 
energy of $\varepsilon_b\simeq 0.5-2.5$ eV where a large 
gap with vanishing density of states appears in our 
calculated results (see Fig.~3).  This result suggests 
the importance of electron correlations in this material; 
as has been pointed out in Ref.~\cite{PES}, the situation is 
similar to the case of V$_2$O$_3$, for which the single-particle 
excitation spectra have been studied in detail.\cite{mo}  
The large and broad spectral weight appears at 
$\varepsilon_b\simeq 2.5-9$ eV, which comes mainly from 
the O 2$p$ orbitals.  The position and shape of this 
spectral weight agree well with our calculated density 
of states shown in Fig.~3.  

Below the transition temperature, the coherent peak near 
the Fermi energy has been reported to shift largely to higher 
binding energies, reflecting the opening of the quasiparticle 
gap of 230 meV.\cite{PES}  
The reconstruction of the electronic states by the 
metal-insulator transition should be pursued theoretically.  
However, the LDA and GGA calculations cannot reproduce 
the charge-ordered states with the spin-singlet formation.  
One needs to invent the strongly-correlated electron models 
to account for the phase transition, of which the simplest 
one-dimensional version has been proposed in 
Ref.\cite{horiuchi}.  Further study is required to take into 
account of the three-dimensionality of the system.  

It has been reported in a recent NMR experiment\cite{NMR} 
that there appears an exceptionally strong temperature 
dependence of the rotation of the NMR Knight-shift tensor, 
which suggests the strong orbital dependence of the local 
spin susceptibility.  The degeneracy of the $d_{yz}$ and 
$d_{zx}$ orbitals may play a role here.  
We point out that the essential information should be obtained 
if the calculation of the spin density distribution under a 
uniform magnetic field can be made, which we leave for 
our future study.  

It has also been reported\cite{yamauchi} that, under high 
pressures, a variety of charge-ordered phases appear in this 
system; in particular, a different charge-ordered pattern 
with the antiferromagnetic long-range order has been 
suggested to occur above $\sim$1 GPa at low temperatures.  
Competing phases with similar energies have therefore been 
suggested to exist in this system, which should further 
be clarified both experimentally and theoretically.

\section{Summary}

We have carried out the \textit{ab initio} electronic 
structure calculation of hollandite vanadate K$_2$V$_8$O$_{16}$ 
to clarify its basic electronic states in the weak correlation 
limit.  
We have shown the following: 
\par\noindent
(i) The states near the Fermi level consist predominantly 
of the three $t_{2g}$ orbitals and the hybridization with 
the oxygen $2p$ orbitals is small.  The strong-coupling model 
based on the three $t_{2g}$ orbitals may therefore be 
justified.  
\par\noindent
(ii) The $d_{yz}$ and $d_{zx}$ orbitals are exactly 
degenerate and are lifted from the $d_{xy}$ orbital.  
The doubly-degenerate atomic energy levels may play a 
role in the observed electronic properties of the system.  
\par\noindent
(iii) The calculated band structure and Fermi surface 
indicate that the system is not purely one-dimensional 
but the coupling between the VO double chains is important.  
\par\noindent
(iv) No nesting features in the Fermi surfaces that contribute 
to the instability of the doubling of the unit cell along 
the $c$ axis are seen, suggesting the observed lattice 
instability to be the strong-coupling origin.  
\par\noindent
(v) The presence of strong electron correlations is 
suggested in the observed valence-band photoemission 
spectra compared with the calculated density of states.  

We hope that the present results obtained in the weak 
correlation limit will be combined with the theory based 
on the strongly-correlated electron models in near future, 
to help us understand the nature of the charge, orbital, 
and spin degrees of freedom of this intriguing material.  

\begin{acknowledgments}
We would like to thank M. Isobe, M. Itoh, Y. Shimizu, 
A. Yamasaki, and T. Yamauchi for useful discussions on 
the experimental aspects of K$_2$V$_8$O$_{16}$.  
This work was supported in part by Grants-in-Aid for 
Scientific Research (Nos.~18028008, 18043006, 18540338, 
and 19014004) from the Ministry of Education, Culture, 
Sports, Science and Technology of Japan.  
A part of computations was carried out at the 
Research Center for Computational Science, 
Okazaki Research Facilities, and the Institute 
for Solid State Physics, University of Tokyo.  
\end{acknowledgments}

\end{document}